\documentclass[journal,compsoc]{IEEEtran}
\renewcommand{\normalsize}{\fontsize{9.5}{11.5}\selectfont}
\usepackage{epsfig,rotating,setspace,latexsym,amsmath,epsf,amssymb,amsfonts,bm,theorem,epstopdf,cite,mathtools,caption,subcaption,enumerate,longtable,accents,bbm,algorithm,graphicx,epsf,authblk,url,color,multirow,comment,tabularx,dsfont,physics,tikz,pifont}
\usepackage{mathtools, cuted}
\usepackage[inline]{enumitem}
\usepackage[noend]{algpseudocode}

\algrenewcommand\algorithmicforall{\textbf{foreach}}
\algrenewcommand\algorithmicindent{.8em}

\newtheorem{theorem}{Theorem}

\newtheorem{definition}{Definition}

\newtheorem{lemma}{Lemma}

\newenvironment{Proof}[1]{\medskip\par\noindent{\bf Proof:\,}\,#1}{{\mbox{\,$\blacksquare$}\par}}

\IEEEoverridecommandlockouts
\allowdisplaybreaks

\begin{document}

\title{SEDULity: A Proof-of-Learning Framework for Distributed and Secure Blockchains with Efficient Useful Work}

\author{Weihang Cao \qquad Mustafa Doger \qquad Sennur Ulukus\\
\normalsize Department of Electrical and Computer Engineering\\
\normalsize University of Maryland, College Park, MD 20742 \\
\normalsize \emph{whcao@umd.edu} \qquad \emph{doger@umd.edu} \qquad \emph{ulukus@umd.edu}}

\maketitle

\begin{abstract}
The security and decentralization of Proof-of-Work (PoW) have been well-tested in existing blockchain systems. However, its tremendous energy waste has raised concerns about sustainability. Proof-of-Useful-Work (PoUW) aims to redirect the meaningless computation to meaningful tasks such as solving machine learning (ML) problems, giving rise to the branch of Proof-of-Learning (PoL). While previous studies have proposed various PoLs, they all, to some degree, suffer from security, decentralization, or efficiency issues. In this paper, we propose a PoL framework that trains ML models efficiently while maintaining blockchain security in a fully distributed manner. We name the framework SEDULity, which stands for a \underline{S}ecure, \underline{E}fficient, \underline{D}istributed, and \underline{U}seful \underline{L}earning-based blockchain system. Specifically, we encode the template block into the training process and design a useful function that is difficult to solve but relatively easy to verify, as a substitute for the PoW puzzle. We show that our framework is distributed, secure, and efficiently trains ML models. We further demonstrate that the proposed PoL framework can be extended to other types of useful work and design an incentive mechanism to incentivize task verification. We show theoretically that a rational miner is incentivized to train fully honestly with well-designed system parameters. Finally, we present simulation results to demonstrate the performance of our framework and validate our analysis.
\end{abstract}


\section{Introduction}
The world has witnessed the substantial potential of blockchain for its capability to reach agreements in a decentralized and trustless environment \cite{Wang2019, Yue2021, Alghamdi2024, Liu2025}. In a blockchain system, nodes maintain data integrity and security by following the adopted consensus protocol, one of blockchain's most essential components. As one of the most representative consensus protocols, Proof-of-Work (PoW) has demonstrated its well-tested security and scalability in various blockchain systems such as Bitcoin \cite{Nakamoto2008}. However, its tremendous and inefficient utilization of energy and computation poses significant sustainability challenges and has been widely criticized \cite{Yu2024}. In PoW, to add new records in the form of blocks, blockchain maintainers, also known as miners, are required to find a solution to a difficult cryptographic puzzle by brute force, a task serving no purpose other than maintaining blockchain consistency.  To maintain a relatively stable block generation interval, as more miners join the network, the puzzle difficulty increases, and the amount of electricity and computation consumption surges considerably. For instance, the annual electricity consumption in Bitcoin is now comparable to that of a medium-sized country such as Poland \cite{Ghadertootoonchi2023}. Energy consumption thus remains a pressing issue that limits the application of blockchain.

Proof-of-Useful-Work (PoUW) aims to address the inefficient energy usage by utilizing useful real-life puzzles as a substitute for the useless PoW puzzle, so that the energy and computation originally wasted in PoW can be repurposed for solving useful tasks, and the blockchain network can reach consensus while doing some useful work simultaneously. Existing studies investigate various useful works, such as finding prime numbers \cite{King2013}, matrix multiplications \cite{Shoker2017, Wei2023, Komargodski2025}, orthogonal vectors problems \cite{Ball2017}, and optimization problems \cite{Shibata2019, Fitzi2022, Todorovic2022, Cao2024}, which are, in general, difficult to solve but relatively easy to verify.

Recently, the rapid development of machine learning (ML) \cite{Liu2020, Ural2023, Yuan2024} has offered a new direction for PoUW. As a special type of optimization problem, ML perfectly meets the requirements of PoUW. In addition, the increasingly fierce global competition in ML has consumed a significant amount of energy and computation, especially with the emergence of large language models (LLMs) \cite{Zhao2023,Minaee2024}, creating a huge demand for ML training. Therefore, Proof-of-Learning (PoL) \cite{Chenli2019, Baldominos2019, BravoMarquez2019, Liu2021, Qu2021, Wang2022, Zhao2024, Cui2025} has been proposed, which combines training ML models with blockchain consensus, and serves the dual function of reaching agreements and solving useful ML tasks simultaneously. 

Despite various PoL proposals, they all, to some extent, face efficiency, security, or decentralization challenges. First, most existing PoLs are based on competitions \cite{Chenli2019, Baldominos2019, BravoMarquez2019, Liu2021, Qu2021, Wang2022, Zhao2024}, where miners compete to solve the same ML task, either being the first or with the highest test accuracy, which is energy-wasting, unfair, and insecure. For each ML task, only the winner is doing unique useful work, while the others' computations and energy are redundant, failing to fully utilize the distributed computing potential of a blockchain network. Moreover, blockchain becomes less secure as only the winner's computing power contributes to the blockchain security. To address security issues, techniques such as submission deadlines \cite{Chenli2019, Liu2021, Wang2022, Cui2025} and validator committees \cite{BravoMarquez2019, Wang2022} are applied. However, these techniques further lead to subjectivity \cite{Daian2019} issues: if a miner joins the network later and is faced with two conflicting records, it is impossible for it to determine which one is authentic without some external trusted sources. PoW, on the other hand, does not have this issue. Therefore, these PoLs can only operate with reduced security assumptions, and are not suitable to be implemented in a fully trustless environment.

To overcome these issues, in this paper, we propose SEDULity, a \underline{S}ecure, \underline{E}fficient, \underline{D}istributed, and \underline{U}seful \underline{L}earning-based blockchain system that achieves both blockchain security and useful work efficiency in a fully distributed manner. This work is based on the same security assumption as PoW, requiring that anyone can verify the chain without any prior trust on any entity. The key idea of this work is that, rather than only allowing the winner of a useful work-based competition to generate a block, each miner is given a block generation opportunity (with some probability) for training a part of the ML tasks. This not only allows miners to work on different ML tasks, but also preserves the security features of PoW against typical attack vectors. Based on this idea, we design a useful function that incorporates ML with blockchain consensus. Notably, our approach can be extended to other kinds of useful work by inserting pseudo-randomness in the computation process (e.g., matrix multiplication). The major contributions of this paper are as follows:

\begin{itemize}
    \item We propose a PoL framework that utilizes ML training for blockchain consensus. Our framework ensures useful work is done efficiently to preserve the security and decentralization of the blockchain.
    \item We demonstrate that our framework can be extended to other types of useful work, using matrix multiplication as an example.
    \item We design an incentive mechanism to incentivize task verification by making minor adjustments to the proposed framework.
    \item We prove that miners are incentivized to train fully honestly with fine-tuned system parameters through rigorous mathematical analysis.
    \item We demonstrate the performance of the proposed PoL through extensive simulations.
\end{itemize}

The rest of this paper is structured as follows. In Section~\ref{sec:related works}, we present related works and discuss the fundamental issues in competition-based PoL. Section~\ref{sec:system model} introduces the system model, and Section~\ref{sec:framework} elaborates on our PoL framework design. \textcolor{black}{We present the security analysis in Section~\ref{sec:security} and provide discussions in Section~\ref{sec:discussions}. Section~\ref{sec:CTF} introduces the capture-the-flag protocol, and Section~\ref{sec:verification analysis} provides the probabilistic verification analysis.}  Simulation results are presented in Section~\ref{sec:sim}. We conclude the paper in Section~\ref{sec:conclusion}.

\section{Related Works} \label{sec:related works}
\subsection{Proof-of-Useful-Work}
The concept of Proof-of-Useful-Work (PoUW) was first formalized and defined in \cite{Ball2017}. In \cite{Ball2017}, it is stated that a PoUW should satisfy the following properties: 
\begin{itemize}
    \item \textbf{Completeness}: A valid proof should be accepted by all honest miners.
    \item \textbf{Soundness}: An invalid proof should be rejected by all honest miners.
    \item \textbf{Hardness}: Generating a valid proof guarantees actual work.
    \item \textbf{Usefulness}: Useful computational tasks are used as challenges, the solution to which can be quickly and verifiably reconstructed from miners' proofs.
    \item \textbf{Efficiency}: Task initializations, proof verifications, and solution reconstructions involve small overhead.
\end{itemize}
Notably, completeness, soundness, and hardness should be satisfied by all public chain consensus mechanisms, while usefulness and efficiency are unique to PoUW. In general, usefulness and efficiency require that the useful work is difficult to solve but relatively easy to verify, just like PoW puzzles. However, we will show in Section~\ref{sec:fundamental problems} that having small overhead for a single miner is not enough for efficiency, but the redundant works from the whole network's perspective should also be small. Additionally, adjustable difficulty is desired to accommodate for the fluctuation of mining power in the blockchain system.

The earliest PoUW can be traced back to Primecoin \cite{King2013}, where finding chains of prime numbers is used as the useful work. Other early works explore tasks such as DNA alignments \cite{Ileri2016}, discrete logarithms \cite{Hastings2018}, traveling salesman problems \cite{Loe2018}, matrix multiplications \cite{Shoker2017}, and computation using trusted hardware \cite{Zhang2017}. However, some of them \cite{Loe2018, Hastings2018} solve randomly generated puzzles that are not truly useful; others \cite{Ileri2016, Shoker2017, Zhang2017} rely on trusted third parties or trusted hardware, which undermines the trustless assumption of the blockchain and induces potential security risks. 

Researchers later realized that optimization problems, omnipresent in engineering, are an important type of useful work. A combinatorial optimization consensus protocol (COCP) solving combinatorial optimization problems is proposed in \cite{Davidovic2022, Todorovic2022}. However, it fails to consider solution theft, where an adversary can directly copy others' solutions, violating the hardness property and threatening blockchain security. Essentially, preventing solution theft is fundamental for all PoUWs, as it is related to both usefulness and security \cite{Cao2024}. Other works design countermeasures against solution theft, but lead to efficiency issues. Shibata \cite{Shibata2019} proposed proof-of-search, which searches for approximate solutions to optimization problems. However, it is far from efficient since optimization problems are treated as black boxes, so that miners can only randomly guess the solutions.  In \cite{Fitzi2022}, a PoUW named Ofelimos is built upon a search algorithm that repeatedly explores the neighborhood to improve the solution. However, the applied pre-hash method not only forbids miners from freely selecting the initial points, which may affect solving performance, but also reduces efficiency. A general optimization-based PoUW framework is formalized and analyzed in \cite{Cao2024}. However, the necessary security overhead hinders the PoUW from being efficient. 

\subsection{Proof-of-Learning}
\subsubsection{Proof-of-Performance}
Machine learning (ML) models are difficult to train but relatively easy to evaluate, making them good candidates for useful work. Inspired by this intuition, most existing PoL proposals focus on verifying the trained models according to their performance (e.g., accuracy) on some test datasets, i.e., Proof-of-Performance (PoP), as noted by \cite{Zhao2024}. The authors of \cite{Chenli2019} proposed Proof-of-Deep-Learning (PoDL), where a new block is valid if and only if a proper deep learning model is produced, and two phases are set up between blocks for model submission. In phase 1, miners train ML models, form candidate blocks containing the necessary block data and the hashed model, and then submit only the block header to full nodes by the end of phase 1. In phase 2, the test dataset is released, and miners submit their models and blocks to full nodes who determine the winner based on the released test dataset. The two-phase submission prevents miners from overfitting the test dataset and stealing others' models, but also from double-spending, since the block headers submitted at the end of phase 1 serve as a commitment of the models. Both overfitting and model theft (which can only occur in phase 2) will make the block header different from what is submitted in phase 1 and invalidate the block. However, this approach requires setting up a checkpoint at every block to finalize the previous records, which has high synchronization demands and is sensitive to network latencies. Consequently, it requires subjectivity assumptions, i.e., reliance upon social information to agree on the current state rather than a fully trustless public blockchain setting as in PoW blockchains. What's more, training competitions lead to redundant training and reduces efficiency.

Instead of two-phase submissions, in \cite{Baldominos2019}, a threshold of minimum performance is utilized to determine the winner. The threshold will gradually decay as time goes by, until a miner is able to derive a qualified model, to maintain reasonable block generation intervals. However, model evaluation is performed by the so-called platform, and many details (e.g., how to manage this test dataset) are unknown. To prevent model theft, in \cite{Baldominos2019}, after a miner prepares a candidate block and computes the block hash, it needs to perform a hash-to-architecture mapping to determine the model architecture used for training; otherwise, the model will not be accepted. However, this method severely impairs model performance and training efficiency. Similarly, in \cite{Liu2021}, the secure mapping layer (SML) is proposed to prevent pre-training and model theft, which is a fixed additional layer before the original input layer determined by the current block and the candidate block. It is experimentally shown that the SML slows down the training process but does not affect the final accuracy. \textcolor{black}{However,} \cite{Liu2021} suffers from the same issues as \cite{Chenli2019} since submission deadlines and ML competitions are still applied. In \cite{BravoMarquez2019, Wang2022}, variations of the Algorand Byzantine agreement (BA) protocol is utilized to select a committee of validators, who are responsible for model ranking and block generation. In this case, ML competitions can be treated as on-chain applications, and the ML training does not directly affect blockchain security. However, Algorand is a voting-based permissioned consensus mechanism using multiple rounds of communications, which is not suitable for public blockchains. Moreover, combining ML with a lightweight consensus mechanism deviates from the original intention of PoUW, i.e., preserving security by making attacks costly.

Given the limited computing power of a single miner, PoL miners may find training a model collaboratively more efficient. \cite{Li2021} demonstrates a mining pool solution for PoDL. Similarly, observing that federated learning (FL) and pooled mining naturally fit in terms of organizational structure, the paper \cite{Qu2021} proposed Proof-of-Federated-Learning (PoFL) that allows miners to form pools and collectively train models. However, a third-party platform is assumed to manage all the federated learning tasks. To tackle this issue, a platform-free Proof-of-Federated-Learning (PF-PoFL) is proposed in \cite{Wang2022}. A Proof-of-Federated-Learning-Subchain (PoFLSC) is proposed in \cite{Li2023} to record and manage the interactions between miners. Federated learning reduces competition and enhances collaborations between miners. However, these FL-based proposals still involve competition among mining pools and suffer from redundant training. \textcolor{black}{Inspired by ensemble learning, the authors of \cite{Cui2025} proposed BagChain for bagging-based decentralized ML, in which base models trained by individual miners can be aggregated into strong ensemble models. However, the complex three-layer blockchain architecture and the three-phase block generation process utilize multiple submission deadlines, which may only be suitable for small-scale blockchain networks with good synchronization. }

Analogous to general PoUWs, model theft is critical in PoP, which is tackled by various methods. However, most of the solutions are far from satisfactory, as they induce new issues related to security and subjectivity. What's more, controllable difficulty remains one of the fundamental obstacles with PoP \cite{Zhao2024}, as it is hard to even evaluate the difficulty of training a model to a certain accuracy, not to mention controlling it, and most existing works in the literature do not recognize or intentionally choose to evade this issue. 

\subsubsection{Proof-of-Computation}
Instead of proving the performance, an emerging line of work in PoL focuses on proving that the exact computation is done for the training, i.e., Proof-of-Computation (PoC) \cite{Zhao2024}. Jia \textit{et. al.} \cite{Jia2021} were the first to point out that the intermediate model weights during training can be used to prove the ownership of the model, i.e., the model is trained in a specific way by someone. In their work, the whole training process is divided into multiple stages, each occupying several epochs. The prover (miner) sends the intermediate model weight of each stage as the proof to the verifier, who then verifies it by recomputing steps with the largest updates. However, later an adversarial example is provided in \cite{Zhang2022a}, which can generate a valid proof with significantly less cost than honest training. \cite{Fang2023} further states that \cite{Zhang2022a} underestimated the lack of robustness in \cite{Jia2021}, and claims that a robust PoL verification mechanism is not possible without further understanding of deep learning.

Based on Jia \textit{et. al.}'s idea, the paper \cite{Zhao2024} proposes probabilistic verification, in which the verified stages are selected at random instead of only the biggest updates, and proposes the capture-the-flag (CTF) protocol to tackle the lazy verifiers who save energy and cost by pretending to verify. The entire work is analytically proven to achieve incentive security, namely, the prover will lose profit if it does not train every stage honestly. However, several designs remain unsatisfactory. First, though it is stated in \cite{Zhao2024} that model theft should be prevented, their PoL actually cannot prevent model plagiarism since the prover's identity is not included in the proof generation. Second, in \cite{Zhao2024}, provers compete to be the first to solve the same ML task, which is energy-wasting, unfair, and insecure. For each ML task, only the winner is doing unique useful work, while the others' computations and energy are redundant, which fails to fully utilize the distributed computing potential of a blockchain network. Moreover, since the prover who first solves the problem wins, if all the provers are honest, the prover with the greatest computing power in the network is supposed to always win. Not only does it induce fairness issues, but other provers may also be incentivized to cheat to increase their chances of winning. In addition, latencies in the blockchain network may make it difficult to determine the winner, leading to diverged views and possibly breaking the consensus. Third, the verification protocol in \cite{Zhao2024} is interactive, requiring several rounds of communication between the prover and the verifier. As a result, the verification time may be subject to network latencies and the prover's bandwidth. PoW, on the contrary, is non-interactive to reduce the negative effect of network latencies on the consensus protocol. Furthermore, though the interactive verification method reduces communication overhead as provers directly send models to the verifiers, subjectivity issues arise as when someone looks back in the future, it is unable to verify these blocks and can only choose to trust the decisions of verifiers. 

To reduce redundant training, a hybrid PoUW protocol based on distributed ML is designed in \cite{Lihu2020}, in which miners collectively train a ML model in a distributed manner by communicating their local model updates \cite{Strom2015}. All the messages during a training task are recorded and supervised by supervisors to prevent malicious behaviors. After each iteration, a miner has the right to mine a block with a limited number of block generation opportunities similar to PoW, whose energy consumption is significantly lower than ML training. Block verifications are delegated to randomly selected verifiers by re-running the lucky iterations. However, this may lead to potential centralization and subjectivity issues since only the supervisors keep the record and only the verifiers verify blocks. To prevent miners from obtaining more block generation opportunities by manipulating the block data, a zero-nonce block must be committed several iterations before. However, this increases confirmation latency and reduces the responsiveness of the blockchain network. Although using block generation opportunities is an insightful idea, a fundamental limitation is that the distributed ML training technique applied in \cite{Lihu2020} has high synchronization demands, which is not suitable for decentralized settings where nodes are asynchronous.

Notably, most existing PoLs operate in a competition-based manner. One of the reasons might be that researchers only notice the similarity between PoW mining competitions and ML competitions, but do not recognize that even in PoW, the mathematical problems solved by different miners are different. However, competition leads to network-scale inefficiency and overhead, even though the overhead of a single miner is acceptable. Indeed, it is stated in \cite{Dikshit2025} that existing PoUWs do not possess substantial energy efficiency, as most of the computing power is still consumed redundantly. \cite{Ural2023} pointed out that competitions in PoLs may further induce scalability challenges. In Section~\ref{sec:fundamental problems}, we will further discuss issues regarding the competition-based consensus systems and summarize the lessons learned.

\begin{table*}[t]
    \centering
    \caption{Comparison of major PoL proposals.} 

    \newcolumntype{P}[1]{>{\raggedright\arraybackslash}m{#1}}
    \newcolumntype{C}[1]{>{\centering\arraybackslash}m{#1}}
    \resizebox{\textwidth}{!}{
    \begin{tabular}{|c|c|C{2cm}|C{1.6cm}|C{3.5cm}|C{4.5cm}|c|c|c|}
    \hline
    
       Ref.& Year & Model Evaluation & Block Creator & Core Design & Major Limitations & Efficiency & Security & Decentralization \\ \hline \hline
       \cite{Chenli2019}  & 2019 & Performance & Winner & Two-phase submission & Redundant training; subjectivity issues & \textcolor{black}{Low}& \textcolor{black}{Medium} & \textcolor{black}{Medium}\\  \hline
       \cite{Baldominos2019} & 2019 & Performance & Winner & Hash-to-architecture mapping & Reduced training performance; redundant training  &  \textcolor{black}{Low} & \textcolor{black}{Medium} &  \textcolor{black}{Medium}\\  \hline
       \cite{BravoMarquez2019} & 2019 & Performance & Validator committee & Proof-of-Storage-based Algorand & Redundant training; subjectivity issues & \textcolor{black}{Low} & \textcolor{black}{High} & \textcolor{black}{Low} \\ 
       \hline
       \cite{Lihu2020} & 2020 & Computation & Lucky miner & Block generation opportunities for training & Subjectivity issues; poor responsiveness; high synchronization demand &\textcolor{black}{High} & \textcolor{black}{Medium} & \textcolor{black}{Low}\\        
       \hline
       \cite{Liu2021}  & 2021 & Performance & Winner & Secure mapping layer & Redundant training; subjectivity issues; prolonged training time & \textcolor{black}{Low} & \textcolor{black}{High} & \textcolor{black}{Medium} \\ 
       \hline
       \cite{Qu2021}  & 2021 & Performance & Winner & Federated learning; privacy-preserving model verification & Trusted platform reliance; subjectivity issues; redundant training & \textcolor{black}{Medium} & \textcolor{black}{High} & \textcolor{black}{Low}\\ 
       \hline
       \cite{Wang2022}  & 2022 & Performance & Validator committee & Federated learning; credit-based Algorand & Subjectivity issues; redundant training  & \textcolor{black}{Medium} & \textcolor{black}{High} &  \textcolor{black}{Low}\\ 
       \hline
       \cite{Zhao2024}  & 2024 & Computation & Winner & Encoding block data into training & Cannot prevent model theft; redundant training; fairness and subjectivity issues  & \textcolor{black}{Low} & \textcolor{black}{Low} &  \textcolor{black}{Medium} \\ \hline       
       \textcolor{black}{\cite{Cui2025}}  & \textcolor{black}{2025} & \textcolor{black}{Performance} & \textcolor{black}{Lucky miner} & \textcolor{black}{Ensemble learning} & \textcolor{black}{Redundant training; high synchronization demand}  & \textcolor{black}{Medium} & \textcolor{black}{High} &  \textcolor{black}{Medium} \\ \hline 
       Ours & 2025 & Computation & Lucky miner & Encoding block data into training & Above limitations are overcome  & \textcolor{black}{High} & \textcolor{black}{High} & \textcolor{black}{High}\\ 
    \hline
    \end{tabular}}
    \label{tab:comparison}
\end{table*}

\subsection{Fundamental Problems of Competition-based PoL -- Lessons Learned} \label{sec:fundamental problems}
In this subsection, we demonstrate the fundamental problems of competition-based PoL frameworks using the following representative examples. We start with a PoW blockchain network of 5 honest miners, each occupying $20$ units of the total mining power of $100$ units. In PoW, miners attempt to generate a new block by repeatedly feeding randomly selected numbers, also known as nonces, into a hash function and observing whether the output is smaller than a given threshold, a process similar to drawing a lottery. Since all miners have the same computing power, the number of nonces they can try per unit time is the same. Thus, each miner has the same probability of generating a new block. If an adversary wants to launch a double-spending attack by generating a longer fork to exclude a target block from the main chain, it needs to generate blocks faster than the main chain. This means that, in an ideal network with no latency, the adversary needs to control more than $100$ units of additional computing power to make sure its double-spending attack is successful.

Now, replace PoW in the previous example with a competition-based PoL where the first miner who solves a given ML task can generate the next block. This is still a fair competition as each miner has the same probability of winning. However, since all miners are solving the same problem, only one of their solutions is useful, i.e., only $20$ units of the mining power contribute to the useful work. The useful work efficiency would further decrease with more miners. This means that, although competition-based PoL solves useful problems, it is still highly energy-wasting.

Then, suppose that an adversary aims to launch a double-spending attack. Note that the block generation interval is the time it takes to solve an ML task using $20$ units of mining power. Thus, if the adversary controls more than $20$ units of mining power, it can generate blocks faster than the growth rate of the main chain. That is, the non-winners' work does not contribute to the blockchain security.

Next, consider another example with 5 honest miners, but miner $1$ possesses $40$ units of the mining power, while each of the other miners has $15$ units of the total $100$ units. In a PoW system, each miner generates a block proportional to the fraction of the computational power it has in the system. In a competition-based PoL system, since miner $1$ possesses the most computing power, it will almost always be the winner unless some other miner cheats. This is not only unfair but also insecure. 

To mitigate ``biggest takes all'' and enhance security, various techniques are applied such as two-phase submissions \cite{Chenli2019}, introducing federated learning \cite{Qu2021, Wang2022,Li2023}, etc. However, most of them are far from satisfactory: two-phase submissions lead to decentralization issues; federated learning reduces but never eliminates competition. Moreover, to accommodate for competitions, most PoLs are complex and can only operate with subjectivity assumptions, requiring prior trust and synchronization demands. In that sense, they offer little advantage in sustainability, security or decentralization compared with lightweight consensus mechanisms such as Proof-of-Stake (PoS).

To conclude, the following \emph{lessons can be learned} regarding how to design a PoL that is simultaneously secure, efficient, and decentralized:
\begin{itemize}
    \item Model theft and reuse should be prevented, but the countermeasures should not influence efficiency and should be amenable to be applied in a decentralized and trustless manner.
    \item PoL should be conducted in a non-competing manner, e.g., every miner should work on a different training task, or all miners should train a model collectively.
    \item Anyone should be able to verify the blockchain without any external trusted entities, even long after the blocks have been created.
\end{itemize}

Therefore, in our proposed framework, we encode block data into the training process to prevent model theft and reuse. We let every miner work on a different task, so that their computing powers can all contribute to the blockchain security. We guarantee that everyone can verify existing blocks independently without having to trust anyone. Table \ref{tab:comparison} further compares and clarifies the differences between existing works and this work.

\section{System Model \label{sec:system model} }
\subsection{Blockchain Model} \label{sec:blockchain model}
The blockchain is a linked list data structure denoted as $\mathcal{BC}=\left\{ B_{0}\leftarrow B_{1}\leftarrow \ldots \leftarrow B_{h}\right\} $ in which $B_{i}$ is the block at height $i$. We assume a cryptographic hash function (e.g., SHA256), $f_H$, which takes a bit string of any length as input and outputs a binary string of fixed length $L$. The hash function applies a sophisticated mapping such that: 
\begin{itemize}
    \item The resulting values are uniformly distributed over $0,1,2,\ldots,2^L-1$.
    \item It is impossible to retrieve the original input from the output.
    \item The output changes significantly even if the input changes slightly.
\end{itemize}
The hash function establishes the foundation of blockchain security. In this work, we refer to $f_H(\text{data})$ as the summary of the data. A block $B_h$ can then be abstracted by
\begin{align}
    B_h=\{f_H(B_{h-1}),ID,ledger, \pi\},
\end{align}
where $f_H(B_{h-1})$ is a hash pointer referring to the previous block, $ID$ represents the identity of the miner who generated the block (e.g., the miner's address) for receiving block rewards, the ledger contains block data in the form of transactions and smart contracts, and $\pi$ is the proof that makes the block valid and accepted by the entire network. In PoW, miners try to solve a difficult puzzle, i.e., try to find a binary string $\pi$, also called a nonce, such that it satisfies
\begin{align} 
    f_H(B_h)<T, \label{eq:pow}
\end{align}
for a given threshold $T$. Thanks to the properties of hash functions, miners can only exhaustively search for a valid nonce, where each inquiry has a success probability $p\approx T/2^L\ll1$. However, \eqref{eq:pow} is useless other than maintaining blockchain security. Our goal in this paper is thus to develop an alternative function for \eqref{eq:pow} that involves useful work, which we call $f_U$. 

In this paper, we assume that all miners apply the longest chain protocol. That is, when there are multiple valid chains, also called a fork, miners always choose to extend the longest among them. If there are more than one longest chain, e.g., two blocks are generated at the same height, miners choose the first one they have seen to extend, until one becomes longer than the others. The longest chain protocol resolves forks in the blockchain and enforces eventual convergence to a unique ledger. Our results are easily extendable to other forms of consensus protocols. 

We consider three main types of entities in the blockchain network: requesters, provers, and verifiers. Requesters are users who commission ML tasks via blockchain. They supply ML tasks that call for trusted model training with rewards. Since the ML datasets and models are too large to be stored directly on-chain, in this work, we assume that they are stored in some layer-2 distributed storage, e.g., the interplanetary file system (IPFS) \cite{Benet2014}. Block maintainers, or miners, are nodes responsible for determining the ordering of records, which are divided into provers and verifiers. Provers train the ML models, generate proofs to claim rewards, and in the meantime, generate blocks to update the blockchain. Verifiers check whether the proofs generated by the provers are valid. Apart from these three entities, there are also blockchain users who want their transactions or smart contracts recorded on-chain. Notably, a requester is a special type of blockchain user. 

\subsection{Adversary Model}
{\color{black} Assume that there are $n$ miners in the peer-to-peer blockchain network, each representing an equal share of computing power. This can be viewed as a discretization of the total computational power, where a powerful miner corresponds to controlling multiple such miners. We assume that there exists a single adversary controlling $\beta<\frac{1}{2}$ of the computing power, which, according to the above abstraction, corresponds to controlling fewer than half of the miners. The adversary may also control a subset of the requesters, but they can only upload ML tasks in the format defined by the blockchain system. (We will further discuss ML tasks in the next subsection.) Honest miners always abide by the protocol, while the adversary may arbitrarily deviate from the protocol. In a typical PoL system, the following misbehavior or attacks should be taken into account.

\textbf{Attacks on the ML layer.} The adversary may not properly train the ML model as required by the system. It may steal a model from an honest miner and claim that it has trained it. Even further, it may upload a ML task with a pre-trained model via a corrupted requester, then submit the known model to claim the reward. In addition, it may attempt to forge a proof that passes verification with significantly less computing power than actually training the ML model.

\textbf{Attacks on the consensus layer.} The adversary may attempt to tamper with the existing blockchain data, which is a fundamental threat considered by all blockchain systems. Specifically, the adversary secretly may mine an alternative chain, and once the secret chain exceeds the main chain in length, the adversary may reveal it to alter the records in the main chain. Moreover, in an anonymous system like the blockchain, the adversary may create multiple identities to gain a potential advantage in mining or possible attacks.

Security against specific attacks will be discussed in Section~\ref{sec:security}. In this work, we assume that the adversary cannot delay or manipulate messages by the honest nodes.

}

\subsection{Machine Learning (ML) Model \label{subsec: ml training}}
An ML task is defined as $A=\left(D_{tr},\mathcal{E},\phi\right)$ in which $D_{tr}$ is the training dataset, $\mathcal{E}$ is the set of environmental variables, and $\phi$ is the random seed. The set $\mathcal{E}$ contains the initial model weight $w_{0}$, the learning rate $\eta$, the loss function $\mathcal{L}$, the batch size $b$, the number of epochs $E$, and the randomization guideline $\mathcal{G}$.

The ML task requires training the model using the stochastic gradient descent (SGD) algorithm for $E$ epochs, each corresponding to a full pass of the training dataset. At the start of each epoch, the dataset is randomly shuffled to avoid bias, and we denote the shuffle function $\sigma_\phi(D_{tr})$ that shuffles the dataset according to the random seed $\phi$. The shuffled dataset is then divided into $b$ batches, which are fed into the model sequentially for training. Thus, the model weight at epoch $e$, denoted as $w_{e}$, can be written as a function of the weight at the previous epoch, $w_{e-1}$, and the shuffled dataset, $\sigma_\phi(D_{tr})$,
\begin{align}
    w_{e}=f_M\left(w_{e-1},\sigma_{\phi}\left(D_{tr}\right)\right). \label{eq:we}
\end{align}
Here, $w_e$ is unique given $w_{e-1}$ and $\sigma_\phi(D_{tr})$, thus can be reproduced by performing \eqref{eq:we}. \footnote{\textcolor{black}{The noise arising from low-level components \cite{Fang2023} can be eliminated by requiring miners to use unified software versions or by reaching a consensus on the most significant bits. Thus, noise-resistant training reproduction can be safely assumed.}}

\section{PoL Framework \label{sec:framework}}
\subsection{Security Deposit and Group Appointment}

\begin{figure*}[t]
	\centering
	\includegraphics[width=\textwidth]{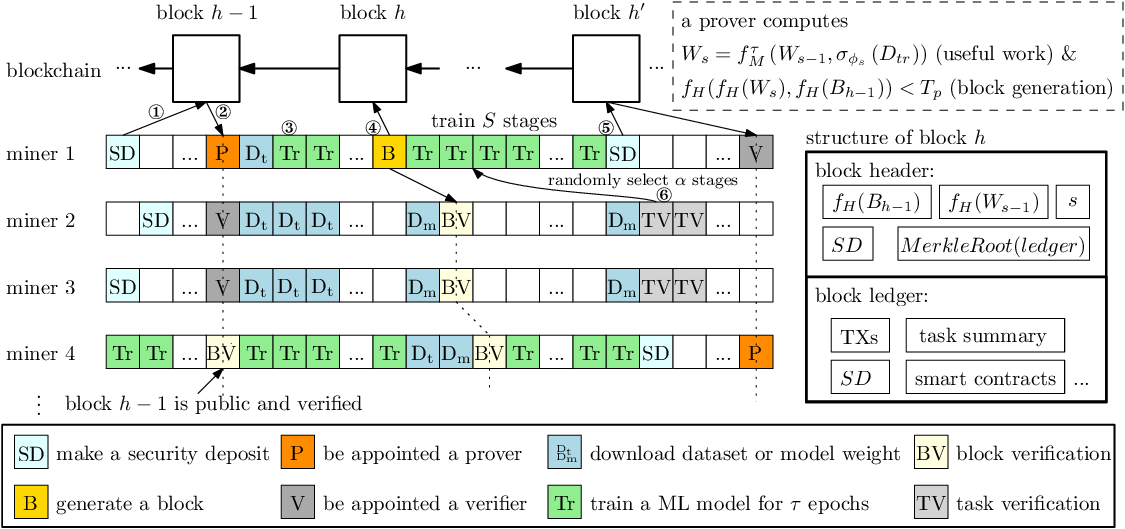}
	\caption{The workflow of the proposed PoL framework, where each small box represents a time-step during which miners complete an action, such as training a stage, verifying a block or downloading some data. \ding{172}: Miners upload their security deposits on-chain and get appointed as provers or verifiers. \ding{173}: After task assignment, provers and verifiers download the corresponding datasets. \ding{174}: A prover (miner $1$) prepares a template block and trains ML models to check whether the template block is valid. \ding{175}: When a block is generated, all other miners take a break from their work and verify the new block. \ding{176}: After the prover (miner $1$) finishes $S$ stages of training, it uploads a new security deposit to get a new assignment. \ding{177}: Verifiers (miners $2$ and $3$) in the same group as miner $1$ randomly select $\alpha$ stages for task verification.}
	\label{fig:fig1}
\end{figure*}

To participate in the consensus process, a miner first locks some of its credit away in a smart contract as the security deposit, denoted as $SD$, which is packaged into a later block and added to the blockchain. The security deposit is related to the miner's address $addr$, i.e., $SD=SD(addr)$, and thus represents the miner's identity. It allows the blockchain network to easily punish misbehaving miners economically. If misbehavior is detected, e.g., the miner is caught cheating in generating a proof, the credit in the security deposit will be deducted. Conversely, a miner can claim to retrieve the credit in the security deposit at any time, and it will be refunded after several blocks if no misbehavior is detected. 

The security deposit also serves as a certificate for appointing provers and verifiers. To guarantee that tasks can be efficiently solved and verified, miners cannot choose to be provers or verifiers. Instead, their roles are randomly assigned. Suppose the security deposit $SD$ is recorded in block $B_h$ and uploaded on-chain. After verifying the block, miners first compute \textcolor{black}{the hash of the security deposit, $h_{SD}$}
\begin{align}
    h_{SD}=f_H\left(f_H\left(B_{h}\right),SD\right), \label{eq:h_sd}
\end{align}
\textcolor{black}{for all new security deposits, where $f_H$ is the hash function defined in Section~\ref{sec:blockchain model}. After the block is received, a uniform record of unassigned $SD$s is maintained by each miner, which is sorted in ascending order. The $SD$s are then divided into groups of size $g$ according to their sorted order. Note that the hash of block $h$, $f_H\left(B_{h}\right)$, is included to prevent miners from knowing the order before the block is generated.}  If the last group has fewer than $g$ miners, this last group of miners will wait in the buffer for the arrival of new security deposits in the subsequent blocks until the buffer is filled. Within each group of size $g$, $g_v$ miners with the smallest $h_{SD}$'s are appointed as verifiers, and the rest $g_p=g-g_v$ miners are appointed as provers. Each prover will be assigned a task (Section~\ref{sec:task assignment}), and each verifier is responsible for validating the ML training (Section~\ref{sec:task verification}). \textcolor{black}{Of course, all miners are responsible for the validation of new blocks generated by provers (Section~\ref{sec:task solving})}.
The group appointment guarantees that each prover's ML training will be verified by a committee of $g_v$ verifiers, and each verifier has the same workload of verifying $g_p$ provers.

\subsection{Task Assignment} \label{sec:task assignment}
Requesters upload their ML tasks to the layer-2 storage. For each task, a smart contract $SC$ containing the summary of the task needs to be recorded on-chain, and the requester needs to lock some of its credits in this smart contract as the task reward to compensate for the cost of the provers and verifiers. Locking the task reward in advance helps prevent malicious requesters from flooding the system with meaningless tasks. For a certain blockchain record, all unassigned ML tasks are fixed and thus form a task pool. Assume that there are in total $P$ tasks in the task pool.

We would like different provers to work on different problems as much as possible. Therefore, provers cannot choose which problem to solve but are assigned tasks randomly. Specifically, similar to group appointments, miners compute \textcolor{black}{the hash of the task smart contract, $h_{SC}$,}
\begin{align}
    h_{SC} = f_H(f_H(B_h),SC),
\end{align}
\textcolor{black}{for each task in the task pool, and maintains a list of unassigned on-chain tasks sorted in ascending order.} Then, starting from the prover with the smallest $h_{SD}$ and the task with the smallest $h_{SC}$, provers are assigned tasks one by one, until there are no more unassigned provers or tasks. If tasks are insufficient, provers wait for the arrival of new blocks.

\subsection{Incorporating Machine Learning into Block Generation} \label{sec:task solving}
After task assignment, provers and verifiers start to download the corresponding tasks, including the datasets. After downloading, a prover starts to train the ML model for $E$ epochs. The $E$-epoch training is divided into $S$ stages, each occupying $\tau=E/S$ epochs. Denote $W_{s}=w_{s\tau}$ as the intermediate model weight after training the stage $s\in\left\{ 1,2,\ldots,S\right\}$. Before training the stage $s$, the prover sets $\pi=\{f_H(W_{s-1}),s\}$ and prepares a potential block $B_h$ at height $h$  (assume the current block height is $h-1$). 
Here, the security deposit is used as the miner's identity, thus
\begin{align}
    B_h=\{f_H(B_{h-1}),SD,ledger, f_H(W_{s-1}),s\}.\label{eq:block_structure}
\end{align}
{\color{black}Next, we explain how block generation opportunities (BGO) give the prover a chance for each training stage to make this potential block valid.} For the training of stage $s$, {\color{black}the prover first computes the random seed $\phi_s$, which} is generated deterministically using the summary of the potential block at height $h$, $f_H(B_{h})$, and the stage number $s$
\begin{align}
\phi_s=g\left(f_H(B_{h}),s\right), \label{eq:phi}
\end{align}
in which $g\left(\cdot\right)$ is some deterministic function that can be reproduced. {\color{black}After obtaining the random seed, the prover trains the model at stage $s$ using \eqref{eq:we} to obtain $W_{s}$ as}
\begin{align}
    W_s=f^\tau_M\left(W_{s-1},\sigma_{\phi_s}\left(D_{tr}\right)\right) \label{eq:Ws}
\end{align}
where $f_M^{\tau}\left(\cdot\right)$ denotes training the model for $\tau$ epochs. After training stage $s$ and obtaining $W_s$, the prover checks whether
\begin{align}
    f_H(f_H(W_s),f_H(B_{h-1}))<T_{p}, \label{eq:BGO}
\end{align}
where $T_{p}=p\cdot2^{L}$ is a given threshold for block generation. If \eqref{eq:BGO} is satisfied, $B_h$ is a valid block. 

The prover uploads $W_{s-1}$ and $f_H(W_s)$ to the layer-2 storage and broadcasts $B_h$ to the blockchain network. To verify $B_h$, all miners, i.e., all provers and verifiers in the system no matter which group they are assigned to, first download the summary $f_H(W_s)$ and check whether \eqref{eq:BGO} holds. If so, they download $B_h$, $D_{tr}$ and $W_{s-1}$ to check the validity of $f_H(f_M^\tau\left(W_{s-1},\sigma_{\phi_s}\left(D_{tr}\right)\right))=f_H(W_s)$ as well as if the block ledger is semantically correct with respect to the blockchain history. If all conditions are satisfied, the block is accepted.

Combining \eqref{eq:Ws} and \eqref{eq:BGO}, we have designed a function
\begin{align}
    f_U=f_H(f_H(f_M^\tau\left(W_{s-1},\sigma_{\phi_s}\left(D_{tr}\right)\right)),f_H(B_{h-1})), \label{eq:fu}
\end{align}
for the PoL framework which contains a useful work $f_M^\tau$ and hence $f_U<T_p$ as the useful problem. In each stage of training, the provers create a potential block with new contents to accommodate received messages, e.g., new transactions or new blocks. However, once a valid block is generated, the block can no longer be tampered with, as doing so changes the random seed $\phi$, making the proof invalid.

The prover who generates a valid block will get a block reward. We would like the provers to finish training all stages of their tasks regardless of whether a block is generated at an intermediate stage. Therefore, the block reward is withheld until the prover finishes the whole task and passes the task verification, {\color{black}which can be easily imposed via smart contracts}. Notably, as long as the stage that generated a block is trained honestly, even if the rest of the training stages were dishonest, {\color{black} the blockchain growth process is not disrupted since the block remains valid}. Hence, the block verification, i.e., verifying \eqref{eq:fu}, which is done by every miner, is independent of the task verification. Task verification is done by the committee of $g_v$ verifiers in the assigned group when the prover completes all $S$ stages of the training and uploads all the intermediate weights. We will discuss the task verification in more detail in the next subsection.

\subsection{Task Verification} \label{sec:task verification}
After training stage $s$, the prover uploads a proof package $\mathcal{P}_s=\{W_{s-1},\phi_s,f_H(W_s)\}$ to the layer-2 storage. When all the stages are completed and all the proof packages are uploaded to the layer-2 storage, the prover uploads the final weight $W_S$ and anyone can verify the proofs by reproducing the computation. However, since verifying all stages by everyone is costly, only the committee of $g_v$ verifiers in the same group is responsible for the task verification, and instead of verifying all stages, each verifier randomly selects $\alpha$ out of $S$ stages to verify. Note that, the verifiers can start downloading $D_{tr}$ right after their groups are determined, hence by the time a task is completed and task verification starts, a verifier already has $D_{tr}$. To verify a selected stage $s$, the verifier requests $\mathcal{P}_s$ from the platform, and checks whether $f_H(W_{s})=f_H(f^\tau_M\left(W_{s-1},\sigma_{\phi_s}\left(D_{tr}\right)\right))$.  A task proof passes verification if all $\alpha$ selected stages of all $g_v$ verifiers are valid. If any of the $g_v$ verifiers finds a stage trained dishonestly, it shares this information with other verifiers in the group, who will also verify the same stage and agree that it is not trained honestly, and the proof will not pass. An example workflow of the proposed PoL framework is given in Fig.~\ref{fig:fig1}.

\subsection{Task Finalization, Rewards, and Punishments}
After verification, the verifiers collectively generate a verification smart contract and upload it to the blockchain to finalize the task and enforce rewards and punishments. Once the verification smart contract is recorded on-chain, the task reward will be distributed among the prover and verifiers. The prover receives a task reward if it provides a valid task proof; however, if cheating is detected, the prover obtains no reward, and the security deposit in $SD$ will be forfeited. Each verifier also obtains a reward to compensate for their verification cost, regardless of whether the task proof is valid. To guarantee task finalization, if the prover fails to generate a proof in a long time, the verifier committee has the right to finalize the task, and the prover will face some penalty.
 
When a task is finalized, the prover can generate a new security deposit and wait for a new block to start a new turn of group appointment. A verifier can do the same once it finishes verifications for all assigned provers of its group. To prevent computing power decrease, when the prover is waiting for the next block, it can re-train any stage $s$ using the previously computed $W_{s-1}$, in order to mine a new block. We will discuss this in detail in Section~\ref{sec:tradeoff}.

{\color{black}
\section{Security Analysis} \label{sec:security}
In this section, we discuss how the proposed PoL framework performs against possible attacks. We assume that computing a single hash summary $f_H(x)$ for an input $x$ is trivial and can be completed within a negligible time. Similarly, downloading a summary $f_H(x)$ is also trivial. On the other hand, neither downloading a model weight $W$, a training dataset $D_{tr}$, nor computing $f_M^\tau(W)$ for an input $W$ takes negligible time.

\subsection{Security Against ML-layer Attacks}

Let us consider a newly generated (maybe invalid) block by the adversary, who uploads $W_{s-1}$ and $f_H(W_s)$ to the layer-2 storage and broadcasts $B_h$ in the network. All the miners in the network take a break from their own work to check the validity of the block (as in Nakamoto consensus). Instead of downloading the entire block and model weights $W_{s-1}$ immediately, the miners first download the summary $f_H(W_s)$ (they already have $B_{h-1}$ in storage) and check \eqref{eq:BGO}, which is trivial and consumes negligible time. Since a random $f_H(W_s)$ satisfies \eqref{eq:BGO} with probability $p\ll1$, even if the adversary generates some random $f_H(W_s)$ instead of doing the ML training, to convince honest miners to stop their own work and download non-trivial data $B_h$, $D_{tr}$ and $W_{s-1}$, the adversary needs to do some non-trivial search to make sure that $f_H(W_s)$ satisfies \eqref{eq:BGO}. Furthermore, $f_H(B_{h-1})$ in \eqref{eq:BGO} ensures that the same $f_H(W_s)$ cannot be used for different block heights. If \eqref{eq:BGO} is not satisfied, the adversary can be immediately punished by losing credits in its $SD$. 

Once miners confirm that \eqref{eq:BGO} is satisfied, they enter the second stage of block verification where they download the non-trivial data $B_h$, $D_{tr}$ and $W_{s-1}$ and semantically validate the block $B_h$, such as checking the ledger with respect to history, recomputing and validating the summary field $f(W_{s-1})$, etc. Finally, they obtain $\phi_{s}$ from \eqref{eq:phi} and validate $f_H(f_M^\tau\left(W_{s-1},\sigma_{\phi_s}\left(D_{tr}\right)\right))=f_H(W_s)$ which involves the non-trivial function $f_M^\tau$. If any of the validation steps fails, the adversary will be punished.

The above analysis indicates that the proposed framework is secure against ML-layer attacks. First, if the adversary generates some random $f_H(W_s)$ that satisfies \eqref{eq:BGO} (with some non-trivial work), finding a valid $W_{s-1}$ that matches $f_H(W_s)$ still requires training the ML model, as the hash of the updated model cannot be obtained without computing $f_M^\tau\left(W_{s-1},\sigma_{\phi_s}\left(D_{tr}\right)\right)$. Therefore, forging a valid proof has the same (or even more) difficulty as training honestly. Second, according to \eqref{eq:block_structure}-\eqref{eq:Ws}, the trained ML model is sensitive to the random seed $\phi_s$, which is determined by the block content containing the previous block hash $f_H(B_{h-1})$ and the miner's identity $SD$. As a result, if the adversary steals a model from an honest miner, the derived updated model $W_s'$ will be different due to a different random seed $\phi_s'$, and the block will become invalid. (To obtain the block reward, the adversary must change the $SD$ to its own.) Third, by the same argument, having a pre-trained ML model does not give the adversary any advantage to generate the next block $B_h$, as the validity of the block depends on the previous block $B_{h-1}$, changing which alters the random seed and invalidates the training.

\subsection{Security Against Consensus-layer Attacks.}

In the proposed framework, $W_{s-1}$ essentially serves the purpose of a nonce in PoW. However, unlike PoW, generating a block involves some non-trivial ML training, i.e., the useful work. Given the current block $B_h$, we can easily see that tampering with any block $B_i$ with $i\leq h$, e.g., changing the identity of the prover that generated the block, invalidates all subsequent blocks, as each block has a summary of the previous block. Hence, the proposed framework is essentially a resource-based blockchain system, where the adversary's ability to rewrite the blockchain history is fundamentally determined by its share of computational resources. 

Suppose the adversary, in possession of $\beta$ fraction of the total computing power, wants to alter the data in a block by mining a secret chain that exceeds the main chain in length. Denote the average block generation interval of the main chain as $t_b$. For each block mined on the main chain, all honest miners take a break from their work, download the dataset and the model, then train the model for one stage to verify the block. Denote the time it takes to download the dataset, download the model, and train the model for one stage as $t_{dd}$, $t_{dm}$, and $t_{tm}$, respectively, and thus the total time that honest miners are away from ML training is $t_v=t_{dd}+t_{dm}+t_{tm}$. Note that the verification latency is small for PoW, but longer in our framework due to downloading and training. On the other hand, the adversary working on the secret chain can ignore block verification and consistently work on block generation. Moreover, generating multiple identities does not increase the adversary’s effective power in altering the blockchain record.

In a resource-based blockchain system, the condition under which the adversary's success probability vanishes exponentially with respect to the confirmation depth is that the honest miners possess the majority of the computing power. In the proposed PoL framework, due to verification latency, this condition becomes
\begin{align}
    \frac{\beta}{\frac{t_b-t_v}{t_b}(1-\beta)}<1,
\end{align}
i.e.,
\begin{align}
    \beta < \frac{t_b-t_v}{2t_b-t_v}. \label{eq:adv_power}
\end{align}
When $t_v\ll t_b$, \eqref{eq:adv_power} becomes $\beta < 0.5$, which corresponds to the honest majority condition in PoW. When $t_v$ increases, the tolerance to the adversary's computing power ratio decreases, as the adversary's effective computing power ratio increases.

\subsection{Useful Work Security}
There is no central authority or delegation of trust to a committee in our consensus mechanism. As we noted earlier, block verification is independent of the task verification, as everyone validates a block, no matter what happens with the task verification. Hence, the committee of $g_v$ verifiers is not a central authority of the consensus mechanism. If a prover cheats and provides an invalid model weight, everyone validates it, and all honest miners will reject it. If a prover provides a valid model weight $W_{s-1}$ but cheats in the rest of the training stages, it can be punished if caught by the committee of $g_v$ verifiers. However, the block remains valid as cheating in other stages does not invalidate the proof used in \eqref{eq:fu}, which only involves the valid training stage $s$.

Since the ML task is only verified by a selected group of verifiers, the adversary may attempt to corrupt such a verifier committee by strategically introducing a large number of identities within a short time window, especially when there are few newly joining (or rejoining) miners. In this way, the adversary aims to dominate a verification group and allow colluding provers in the same group to claim rewards without performing ML training. However, launching such an attack requires the adversary to lock a substantial amount of security deposits across multiple identities within a short period. Additionally, the structure of \eqref{eq:h_sd} includes the hash of the block $f_H(B_h)$, making it difficult for the adversary to predict the value of $h_{SD}$ or manipulate the group appointment. As a result, the adversary cannot reliably ensure dominance in a verification group without committing a substantial amount of credits, and this type of attack does not provide a consistent advantage over honest participation.

}

\section{Discussions} \label{sec:discussions}

\subsection{Useful Work-related Trade-offs} \label{sec:tradeoff}
Incorporating a non-trivial and time-consuming useful work in block generation as \eqref{eq:fu} comes with \textcolor{black}{trade-offs}. When a new block is mined, all other miners take a break from their own work to validate this work. In PoW systems, this is a negligible break, but not in the proposed PoL framework. For instance, assume that $B_{h-1}$ is the latest block in the blockchain system that is already validated and accepted by everyone. To generate the next block, each prover creates a template block $B_h$ and trains the model to check if the block is valid. However, when a prover finds out that its template block is indeed valid and broadcasts it, the rest of the provers are usually somewhere in the middle of their training. Each of the other provers' template blocks is valid with probability $p$, but only creates a fork at height $h$. 

More specifically, assume an honest prover is training stage $s$ using a template for block height $h$ and currently has computed $\tau/2$ epochs of stage $s$. When someone else publishes a block $B_h$, after taking a break from the training and verifying the newly published block, the honest prover has two choices. It either continues training the remaining $\tau/2$ epochs for the sake of completing its task but forgoes the BGO, or creates a new template (with a new seed) for block height $h+1$ and restarts the training $W_{s-1}$ again, which forgoes the initially completed $\tau/2$ epochs {\color{black}in order to get a new BGO for height $h+1$ from the training stage $s$. Since the second option increases the redundant work and honest miners have to finish the task completely before getting rewarded in either case, it is better to choose the first option and forgo the BGO.} Note that, the adversary may choose the first choice but not forgo the BGO, which creates forks. 

Essentially, this phenomenon is caused by the fact that computing \eqref{eq:fu}, which is approximately linear in $\tau$, takes significantly more time and computational power than computing $f_H(B_h)$ in PoW.
Given a fixed $p$, the larger $\tau$ in \eqref{eq:fu}, the more work towards block generation (hence consensus security) is forgone even though the work itself is still useful from the perspective of the ML training. For example, if we were to give a single opportunity for each task ($\tau=S$), after a block is generated on a new height by a prover, the remaining provers have to finish all stages of their tasks without having any BGO. This is one of the main reasons why our model divides training into stages and gives BGOs for each stage (another main reason is that verifying the block requires more redundant training). To mitigate this phenomenon, we decrease $\tau$, but decreasing $\tau$ increases the number of stages $S$, requiring more storage for ML model weights \footnote{\textcolor{black}{The prover needs to upload $S=E/\tau$ different versions of the model for task verification. However, most of these models are not related to block verification, which only requires storing one model per block. Therefore, we can regularly release the storage that is not related to block generation to reduce the storage cost.}}. Moreover, decreasing $\tau$ for a fixed $p$ also creates more BGOs during network latency and thus forks in a time interval. \textcolor{black}{Therefore, both $\tau$ and $p$ should be decreased accordingly to maintain the block generation interval, which creates another trade-off explained next. }

When a prover finishes its task, it essentially makes a new security deposit $SD$ and waits to get assigned a new task in the next block. However, during this time, the active computing power in the system is reduced as the prover remains idle until the next block. If $p$ is too small, more miners become idle, and the active computing power can drop to zero. In this case, no block can be created as miners cannot get new tasks, and the system fails. This issue can be fixed as follows: when a prover has finished its task and is waiting for a new block, instead of waiting idly, it can use any $W_{s-1}$ from a previously finished stage, to create a template for the next block height, re-train the stage to get $W_s'$ and see if the block is valid. Note that, the new $W'_{s}$ is different from $W_s$ the prover obtained earlier, as the random seeds are different. The goal here is to generate a block, not to train the task. This approach ensures that the active computing power does not decrease, but it leads to redundancy of useful work since a $W_{s}$ was already obtained using a different seed ($W_s'$ can be deemed as not useful since it does not contribute to the final model $W_S$). 

The above discussion implies that, if $p$ is too small, the system is not efficient as miners who finish their tasks and wait to be assigned a new task are doing more redundant training. On the other hand, a larger $p$ not only results in a higher fork rate, but also more block verification for the entire network. Hence, in Section~\ref{sec:metrics}, we define new efficiency metrics in terms of the amount of computations contributing to the blockchain security and ML training, which are further investigated in Section~\ref{sec:sim} for different values of $p$.

In conclusion, \textcolor{black}{the increased block verification time and cost lead to inherent limitations for the framework and imply trade-offs that should be carefully considered.} To reduce the useful work that does not contribute to BGO, $\tau$ should be decreased, which in turn increases forks and storage cost. To reduce forks, we can decrease $p$, which in turn increases the amount of redundant useful work. We display this trade-off and consider different choices in our simulations.

\subsection{Useful Work Metrics}\label{sec:metrics}
In the proposed PoL framework, ML training can be done efficiently. However, there still exist computational overheads such as block and task verifications as well as redundant training of stages to prevent the mining power decrease issue discussed in Section~\ref{sec:tradeoff}. In this subsection, we propose two metrics to evaluate the efficiency of the proposed framework. Denote the amount of computations contributing to the final models $W_S$'s as $c_u$, which is the amount of useful work that no miner in the blockchain network has computed before. Denote $c_r$, $c_{bv}$, and $c_{tv}$ as the amount of computations for redundant training, block verification and task verification, respectively.

First, the useful block generation ratio (UBGR) of provers is defined as the amount of ML training that contributes to the final model $W_S$, compared with the amount of ML training contributed to block generation during its service as a prover,
\begin{align}
    \text{UBGR}=\frac{c_u}{c_u+c_r}.
\end{align}
UBGR represents the ratio of useful work contributing to block generations. Then, we define the useful work ratio (UWR) of the blockchain network as the ratio of the amount of ML training that contributes to the final model $W_S$ compared with all training-related computations in the blockchain network,
\begin{align}
    \text{UWR}=\frac{c_u}{c_u+c_r+c_{bv}+c_{tv}}.
\end{align}
Note that, we only consider computations that involve training ML models, and do not take into account hash computations and dataset/model downloading, as they consume negligible amount of energy compared to model training. In Section~\ref{sec:sim}, we use these metrics to evaluate the efficiency of the proposed PoL framework, and show that our framework is still efficient despite the computational overheads.

\textcolor{black}{Notably, in resource-based consensus mechanisms, it is not fair to directly compare the energy consumption since the computational power of the blockchain network is in general exhausted. In that case, the above proposed metrics are useful for the comparison of the energy usage efficiency. For PoW, both the UBGR and the UWR are 0 because there is no useful work. In a competition-based PoL, both the UBGR and the UWR are in general upper bounded by the computing power ratio of the largest miner in the network. However, according to the security assumptions as well as the decentralization requirements, this computing power share is further limited by the maximum allowed adversarial power ratio, which is typically 1/3 or 1/2. Hence, a trade-off between useful-work and decentralization is introduced by competition: increasing the UWR damages the decentralization of power. In contrast, in our proposed PoL framework, the UWR and decentralization do not limit each other.}

\subsection{Extensions to Other Useful Work} \label{sec:extension}

\textcolor{black}{In this work, the reason that we choose ML as the primary useful work is multifold. First, one can naturally insert pseudo-randomness into the stochastic gradient descent algorithm used for ML training, which is essential for preventing model theft and reuse. Second, ML training has epochs, which naturally relate to the stages in our framework. Third, ML training is of adequate hardness so that the amount of data that needs to be uploaded is reduced. Last but not least, ML is now in tremendous demand, making it an essential choice for PoUW. }

\textcolor{black}{Of course, there are other useful work candidates for PoUW.} In this section, we use matrix multiplication \cite{Komargodski2025} as an example to demonstrate how our consensus protocol can be extended to other types of useful work. Given two arbitrary matrices $X$ and $Y$, the requester wants to compute $ Z=X\cdot Y$ secretly (a prover knows nothing about $X$ and $Y$), efficiently (the overhead compared with directly multiplying the two matrices is low), and with uniform difficulty for any $X$ and $Y$. To prevent leaking $X$ and $Y$ to provers, the input matrices are masked by adding low-rank matrices $E$ and $F$, and the requester uploads $X'=X+E$ and $Y'=Y+F$ instead of $X$ and $Y$. A matrix multiplication (MatMul) task thus requires the prover to compute
\begin{align}
    Z'=X'Y'=(X+E)(Y+F), \label{eq:matmul}
\end{align}
from which the requester can retrieve $Z$ by computing
$Z=Z'-(X\cdot F+E\cdot (Y+F))$, which takes few additional computations since $E$ and $F$ are low-rank. Assume for simplicity that both $X$ and $Y$ are $m\times m$ matrices. One can divide $X'$ and $Y'$ into sub-matrices, 
\begin{align}
    X'=\left[X'_{i,j}\right]_{i,j=1}^{m/r}, \quad
    Y'=\left[Y'_{i,j}\right]_{i,j=1}^{m/r},
\end{align}
where $X'_{i,j},Y'_{i,j},\, i,j=1,2,\ldots,m/r$ are sub-matrices of size $r\times r$. Then, $Z'$ can be computed iteratively using the following formula:
\begin{align}
    {Z'}_{i,j}^{(l)}={Z'}_{i,j}^{(l-1)}+X'_{i,a_l}\cdot Y'_{a_l,j}, \quad l=1,2,\dots,m/r, \label{eq:matmul_iter}
\end{align}
where $(a_1,a_2,\ldots,a_{m/r})$ is any permutation of $\{1,2,\ldots,\frac{m}{r}\}$, $ {Z'}_{i,j}^{(l)}$ is an intermediate sub-matrix at step $l$, and $ {Z'}_{i,j}^{(0)}=O_{r\times r}$ for all $i,j$. We denote ${Z'}^{(l)}=\left[{Z'}_{i,j}^{(l)}\right]_{i,j=1}^{m/r}$, and note that ${Z'}^{(m/r)}=Z'$.

Based on the above procedure, a PoUW can be built by slightly changing the proposed PoL framework as follows. During task assignment, a prover is assigned $S$ MatMul tasks instead of one, and solves them sequentially. For task $s\in\{1,2,\ldots,S\}$ denoted as $t_s$, the prover first sets $\pi=\{f_H(t_s),s\}$ as the summary of the task and prepares a template block $B_h$ as follows,
\begin{align}
    B_h=\{f_H(B_{h-1}),SD,ledger, f_H(t_s),s\}.\label{eq:block_structure_matmul}
\end{align}
It then computes the random seed from \eqref{eq:phi} and derives
\begin{align}
    (a_1,a_2,\ldots,a_{m/r})=\sigma_{\phi_s}\left(\{1,2,\ldots,m/r\}\right)
\end{align}
using the shuffle function. After that, it computes \eqref{eq:matmul_iter}, records all the intermediate results $\mathcal{Z}_s=\left[{Z'}^{(l)}\right]_{l=1}^{m/r}$ and checks whether
\begin{align}
    f_H(f_H(\mathcal{Z}_s),f_H(B_{h-1}))<T_{p}. \label{eq:BGO
    MatMul}
\end{align}
If so, the block is valid, and the prover uploads $\mathcal{Z}_s$ and $f_H(t_s)$ to the layer-2 storage. Notably, even though the final result of a MatMul task is invariant of the template block $B_h$, the record $\mathcal{Z}_s$ is dependent on $B_h$ as it affects the order to add the terms together. Therefore, the block cannot be tampered with, a prover cannot forge the record without computing the task even if it knows $Z'$ in advance, and the same record cannot be used more than once to generate a block.

For task $s$, the prover uploads the proof package
\begin{align}
    \mathcal{P}_s=\{\mathcal{Z}_s,\phi_s, f_H( t_s)\}_{s=1}^{S}
\end{align}
to the layer-2 storage. After all $S$ tasks are finished, the verifiers randomly pick some of the tasks, randomly choose some of the intermediate sub-matrices, and verify whether \eqref{eq:matmul_iter} holds. The task proof passes verification if all the verified steps pass.

The above example shows that the proposed PoL framework can also be extended to other useful works, if we can insert pseudo-randomness in the solution process. This is essential for blockchain security, as it prevents solution plagiarism and using the same solution more than once. If solving the entire task is lengthy (e.g., ML), we can divide the whole computation process into sub-tasks of relatively equal difficulty with respect to time (e.g., training $\tau$ epochs), then give a BGO to each sub-task, as we did in the proposed PoL framework. \textcolor{black}{If a task is less difficult, we can also give a BGO to a single task or even a batch of task. For task verification, probabilistic verification can be utilized to reduce computational overhead.}

A potential issue with the above extension is that, while matrix computation has a computation complexity of $O(m^3)$ (this can be made less if some advanced algorithms are used), downloading such a task has a communication complexity of $O(m^2)$, which may take the prover more time to download than to compute. This poses an implicit constraint on the computation-to-communication complexity of the chosen useful work.

\section{Capture-the-flag: Incentivizing Task Verification} \label{sec:CTF}
As we discussed in Section~\ref{sec:security}, task verification is independent of the blockchain security. \textcolor{black}{However, useful work may be impaired if task verification is not sufficiently incentivized.} For instance, if every prover generates the task proof fully honestly, a verifier of the committee can pretend to verify but simply report \textit{pass} for any task proof it is assigned. This increases its profit by reducing verification cost, but also makes task verification unreliable, which in turn stimulates provers to cheat and reduces the quality of useful work. In this section, we design an incentive mechanism to incentivize task verification by making slight changes to the existing framework, inspired by the capture-the-flag (CTF) protocol proposed in \cite{Zhao2024}.

Before training, a prover first generates $\Omega=\left(\omega_{1},\omega_{2},\ldots,\omega_{S}\right)$ as the flags for each stage.  $\omega_s$ takes value $0$, $1$, $2$ with probability $1-\xi$, $\xi/2$ and $\xi/2$, respectively, where $\xi$ is a given flag ratio. 
Before training stage $s$, the prover sets $\pi=\{f_H(W_{s-1}),s,\omega_s\}$ and prepares the template block $B_h$ as follows
\begin{align}
    B_h=\{f_H(B_{h-1}),SD,ledger, f_H(W_{s-1}),s,\omega_s\}.
\end{align}
Then, for all epochs in stage $s$, the new random seed
\begin{align}
    \phi'_{s}=g\left(\phi_s,\omega_s\right), \label{eq:phi_prime}
\end{align}
is used in computing \eqref{eq:Ws} and \eqref{eq:fu} after obtaining $\phi_s$ from \eqref{eq:phi}. This means that the prover has three different ways of training the same stage.
Same as the original framework, the template block is valid if \eqref{eq:BGO} is satisfied. If a block is generated, the prover uploads the proof package to the layer-2 storage and broadcasts $B_h$ to the blockchain network. Since the flag $\omega_s$ is included in the block, all miners can still verify the block by reproducing the computation without going through the CTF protocol.

The key idea of CTF is that, for the stages that do not generate a block, the prover does not reveal the flags $\Omega$ until task verifications are all completed, so that the committee of $g_v$ verifiers has to find out for themselves which flags are used. After training stage $s$, the prover uploads a proof package $\mathcal{P}_s=\{W_{s-1},\phi_{s},f_H(W_s)\}$ to the layer-2 storage, while $\phi'_s$ is not uploaded. When the prover finishes training the whole task, it uploads $\{W_S, f_H(\Omega)\}$ to finish the task proof. During verification, the verifiers randomly select stages except for those that were used to generate blocks. They are required to report flags used in the verified stages, and are extra rewarded if the correct flag $1$ or $2$ is reported.\footnote{\textcolor{black}{The extra reward comes from the requester, which is locked in the smart contract as part of the task reward.}} To obtain the flag of stage $s$, the verifier first tries to verify with $\omega_s=0$. If verification passes, it reports flag $0$. If verification fails, it randomly selects $\omega_s=1$ or $\omega_s=2$ to verify again. If verification is successful this time, it reports the corresponding flag; otherwise, believing that the cheat rate is low, it reports the other flag. After all verifiers in the same group have reported the flags and the verification smart contract is generated, the prover releases the actual flags $\Omega$, which is recorded in the smart contract, and the verifiers check whether they have found the correct flags. If a verifier fails to report for a long time, it is considered faulty, and the prover will still release $\Omega$.  \textcolor{black}{If a reported flag does not match the revealed flag, the corresponding stage will be verified again using the revealed flag by other verifiers in the group. If the revealed flag is incorrect, the prover is found cheating and will be punished; if correct, it means the verifier is not verifying and will be punished.} 

From the prover's perspective, it cannot forge $\Omega$ since $f_H(\Omega)$ is included in the task proof. Therefore, the cheating detection probability of each verified stage is at least $\frac{1}{2}$, and the probability that the prover passes verification is exponentially decaying for each verified stage that is not honestly trained.\footnote{The detection probability of a verified stage is $1$ if the dishonest prover claims flag $0$ and is $1/2$ if it claims flag $1$ or $2$. However, since the flag ratio $\xi$ is predefined, the dishonest prover would look suspicious if every stage is flagged with $1$ or $2$.} We will show in Section~\ref{sec:verification analysis} that miners are incentivized to train fully honestly with a proper penalty, even though CTF makes verifications less reliable. Also, if the prover does not release $\Omega$, the verification will not pass for sure, so it has no incentive to do so.
From the verifier's perspective, though additional computing power is consumed to verify again with another flag, the verifier will receive an extra reward if the correct flag is captured, which compensates for the additional cost. 

\textcolor{black}{Compared with the original framework, for each training stage, a random flag is added to generate the seed and is included in the proof package. Moreover, an additional round of communication is required, but only to transmit a small data payload, i.e., the actual flags, from the prover to $g_v$ verifiers in the same group. Notably, since task verification is independent of block generation, it is less time-sensitive, and the additional round of communication does not affect blockchain security. For task verifications, with a flag rate of $\xi$, on top of the original $\alpha$ stages, a verifier, on average, needs to train the ML model for $\xi\alpha\ll S$ additional stages, which is a small fraction of training an entire task. Hence, the CTF improves incentive compatibility with small additional complexity, communication, and verification overhead.}

\section{Probabilistic Verification Analysis } \label{sec:verification analysis}
Due to probabilistic verification, the prover may not train all $S$ stages honestly, but pretend to have trained some of them to reduce its cost. This section tries to quantify the conditions under which the prover's profit decreases if it does not honestly train the entire task as demanded. If a rational prover gains less profit by cheating, it would choose to train the ML tasks honestly to maximize its payoff, thereby ensuring the quality of the ML training.

Let us consider the case of one prover and one verifier. The prover trains $\rho S$ stages honestly, where $\rho\in\left[0,1\right]$. Denote the task reward as $R_t$ and the total expected block reward obtained from training $S$ stages honestly as $R_b$. Training the entire task honestly has a computational cost denoted by $C$, and a time consumption of $t_M$. We assume that training a stage dishonestly has neither computational cost nor time consumption, so that the prover trains a task at a cost of $\rho C$ using time $\rho t_{M}$. Given all the cheating strategies that cheat on $1-\rho$ proportion of the whole task, denote the maximum probability that the proof passes verification as $q\left(\rho\right)$, which is monotonically non-decreasing with respect to $\rho$, and $q\left(1\right)=1$. 

If the prover's task proof passes verification, it obtains task reward $R_t$ and block reward $\rho R_b$. However, if it is caught cheating, not only will it fail to claim the task and block reward, but a penalty of $\gamma R_t$ will also be forfeited from the prover's security deposit, where $\gamma$ is the penalty-to-task-reward ratio. The expected payoff of the prover, denoted by $u\left(\rho\right)$, is thus given by
\begin{align}
    u\left(\rho\right) =q\left(\rho\right)\left(R_t+\rho R_b-\rho C\right)+\left(1-q\left(\rho\right)\right)\left(-\gamma R_t-\rho C\right).
\end{align}
The expected payoff per unit time, denoted as $\nu\left(\rho\right)$, is 
\begin{align}
    \nu\left(\rho\right)=\frac{u\left(\rho\right)}{\rho t_{M}}, \quad \rho\in\left(0,1\right).
\end{align}
The honest conditions can then be identified by the following definition.

\begin{definition} (Honest Conditions)\label{def:1}
    Provers are incentivized to train fully honestly under the following conditions:
    \begin{align}
        u\left(1\right) & >0,\label{eq:isc1}\\
        u\left(0\right) & <0,\label{eq:isc2}\\
        \nu\left(\rho\right) & <\nu\left(1\right), \quad \forall\rho\in(0,1).\label{eq:isc3}
    \end{align}
\end{definition}

Here, \eqref{eq:isc1} ensures that a completely honest prover can profit from ML training, which is equivalent to $R_t+R_b-C>0$, \eqref{eq:isc2} states that the prover cannot profit when it is completely dishonest, which can be rearranged as
\begin{align}
    \gamma >\frac{q(0)}{1-q(0)}, \label{eq:isc2eq}
\end{align} 
and \eqref{eq:isc3} guarantees that for any $\rho\in(0,1)$, the prover's profit rate is smaller than that of $\rho=1$, such that its best strategy to maximize its profit is to always train honestly, which is equivalent to 
\begin{align}
    q\left(\rho\right)\left(R_t+\rho R_b-\rho C\right)+\left(1-q\left(\rho\right)\right)\left(-\gamma R_t-\rho C\right) \nonumber \\
    < \left(R_t+R_b-C\right)\rho, \quad \forall\rho\in(0,1). \label{eq:isc3eq}
\end{align}

The following lemma is proven in \cite{Zhao2024} to upper bound the probability of passing verification $q(\rho)$ with respect to the honest training ratio $\rho$.
\begin{lemma} (\cite[Appendix E.3]{Zhao2024}) \label{lem:1}
    We have
    \begin{align}
    q\left(\rho\right)\leq\left(1-\kappa+\kappa\rho\right)^{\alpha}\triangleq k(\rho) \label{eq:qrho_upperbound}
    \end{align}
    in which $\kappa$ is the probability to detect cheating when verifying a dishonestly trained stage, e.g., $\kappa=1$ for the original framework and $\kappa=\frac{1}{2}$ for the CTF.
\end{lemma}

Using \eqref{lem:1}, we derive the sufficient honest condition for the proposed PoL framework.
\begin{theorem} (Sufficient Honest Condition) \label{thm:sufficient}
    When $R_t+R_b-C>0$ and $\alpha\geq2$, miners are incentivized to train fully honestly if the following sufficient condition is satisfied,
    \begin{align}
        \gamma &\geq \frac{(1-\kappa)^\alpha}{1-(1-\kappa)^\alpha}. \label{eq:suffi}
    \end{align}
\end{theorem}

\begin{Proof}
\eqref{eq:isc1} is equivalent to $R_t+R_b-C>0$, which is satisfied by assumption. 
When \eqref{eq:suffi} holds, according to Lemma~\ref{lem:1}, 
\begin{align}
    \gamma &\geq \frac{(1-\kappa)^\alpha}{1-(1-\kappa)^\alpha} \geq\frac{q(0)}{1-q(0)},
\end{align}
i.e., \eqref{eq:isc2} is satisfied.

Next, we show that \eqref{eq:suffi} also incorporates \eqref{eq:isc3}.
From Lemma~\ref{lem:1}, we have
\begin{align}
    q\left(\rho\right) & \left(R_t  +\rho R_b-\rho C\right)+\left(1-q\left(\rho\right)\right)\left(-\gamma R_t-\rho C\right) \nonumber \\
    &\leq k(\rho)\left(R_t+\rho R_b\right)-\left(1-k(\rho)\right)\gamma R_t-\rho C \\
    &\leq k(\rho)R_t+\rho R_b-\left(1-k(\rho)\right)\gamma R_t-\rho C. 
\end{align}
Thus,
\begin{align}
    k(\rho)R_t+\rho R_b-\left(1-k(\rho)\right)\gamma R_t-\rho C  \nonumber\\
     \leq\rho\left(R_t+R_b-C\right), \quad \forall\rho\in(0,1)  \label{eq:32}
\end{align}
is a sufficient condition for \eqref{eq:isc3}. Rearranging \eqref{eq:32} and isolating $\gamma$ yields
\begin{align}
    \gamma\geq\frac{k(\rho)-\rho}{1-k(\rho)}=\frac{\left(1-\kappa+\kappa\rho\right)^{\alpha}-\rho}{1-\left(1-\kappa+\kappa\rho\right)^{\alpha}}, \quad \forall\rho \in(0,1). \label{eq:35}
\end{align}
Denote
\begin{align}
    f(\rho)=\frac{\left(1-\kappa+\kappa\rho\right)^{\alpha}-\rho}{1-\left(1-\kappa+\kappa\rho\right)^{\alpha}}, \quad \forall \rho\in(0,1).
\end{align}
Computing the first-order derivative of $f(\rho)$ yields
\begin{align}
    f'(\rho)=\frac{(1+(\alpha-1)\kappa(1-\rho))\left(1-\kappa+\kappa\rho\right)^{\alpha-1}-1}{\left(1-\left(1-\kappa+\kappa\rho\right)^\alpha\right)^2}.
\end{align}
Let $h(\rho)=(1+(\alpha-1)\kappa(1-\rho))\left(1-\kappa+\kappa\rho\right)^{\alpha-1}-1$, then when $\alpha\geq2$,
\begin{align}
    h'(\rho)=\kappa^2\alpha(\alpha-1)(1-\rho)\left(1-\kappa+\kappa\rho\right)^{\alpha-2}\geq0, \quad \forall 0\leq\rho\leq1.
\end{align}
Therefore, 
\begin{align}
    h(\rho)< h(1)=0, \quad \forall \rho\in(0,1).
\end{align}
This means $f'(\rho)<0, \forall \rho\in(0,1)$, i.e., $f(\rho)$ is monotonically decreasing in $\rho$. Thus,
\begin{align}
    \frac{\left(1-\kappa+\kappa\rho\right)^{\alpha}-\rho}{1-\left(1-\kappa+\kappa\rho\right)^{\alpha}}\leq f(0)=\frac{(1-\kappa)^\alpha}{1-(1-\kappa)^\alpha}.
\end{align}
When \eqref{eq:suffi} holds, we have
\begin{align}
    \gamma \geq \frac{(1-\kappa)^\alpha}{1-(1-\kappa)^\alpha} \geq\frac{\left(1-\kappa+\kappa\rho\right)^{\alpha}-\rho}{1-\left(1-\kappa+\kappa\rho\right)^{\alpha}}, \quad \forall\rho \in(0,1), 
\end{align}
indicating that \eqref{eq:32} is satisfied, which is sufficient to show that \eqref{eq:isc3} is also satisfied. Therefore, \eqref{eq:suffi} is a sufficient condition that provers are incentivized to train fully honestly.
\end{Proof}

Theorem~\ref{thm:sufficient} gives a sufficient condition that rational miners would train ML tasks fully honestly. When $R_b+R_t-C>0$ (honest miners can profit) and $\alpha\geq2$ (the verifier verifies more than $1$ stage), as long as \eqref{eq:suffi} is satisfied, any deviation from being fully honest reduces the miner's profit rate, yielding a lower expected payoff. Thus, miners are incentivized to train all stages honestly, guaranteeing the quality of the useful work. 

Theorem~\ref{thm:sufficient} offers an important guidance to the system design that we can guarantee full honesty by setting up abundant penalties to misbehavior. When $\kappa=1$, i.e., the original framework, \eqref{eq:suffi} becomes $\gamma\geq0$. In this case, full honesty can be guaranteed even with no penalty. When $\kappa=\frac{1}{2}$, i.e., the CTF (in the worst case), \eqref{eq:suffi} is equivalent to 
\begin{align}
    \gamma&\geq \frac{\left(\frac{1}{2}\right)^{\alpha}}{1-\left(\frac{1}{2}\right)^{\alpha}} = \frac{1}{2^\alpha-1}, \label{eq:lb-ctf}
\end{align}
which indicates a trade-off between $\gamma$ and $\alpha$. We can reduce the requirements for the penalty-to-task-reward ratio $\gamma$ exponentially with a larger number of verified stages $\alpha$, or reduce the need of $\alpha$ by increasing $\gamma$. Using a proper penalty, a verifier only needs to verify $\alpha=O(1)$ stages per task to guarantee fully honest training. When there is a committee of $g_v$ verifiers instead of just one, since $\alpha\ll S$, the probability of overlap is negligible, and it is similar to just increasing $\alpha$ to $g_v\alpha$.

One may notice that, the sufficient penalty $\gamma R_t$ only concerns the task reward $R_t$ but not the expected block reward $R_b$ or the computational cost $C$. Indeed, in our framework, if a prover keeps training ML models, then in a given time interval, no matter how many tasks are completed, the expected number of blocks it generates and the computational cost should remain the same. Recall that if a prover is caught cheating, it will lose some of the block rewards for some generated blocks, further discouraging a dishonest strategy. Since we are finding a sufficient condition, it is safe to just consider the payoff of the prover in a given time period. 

\section{Simulations} \label{sec:sim}
In this section, we build a simulation prototype using MATLAB and present simulation results to demonstrate the feasibility and performance of the proposed framework. \textcolor{black}{We simulate a virtual blockchain network with $n=1000$ miners, each having an equal computational power.} During group appointment, $g=25$ miners are grouped together, where a committee of $g_v=5$ verifiers is responsible for the task verification of the $g_p=20$ provers in the group. \textcolor{black}{Provers are assigned tasks that require training an ML model for $\overline{E}=4000$ epochs on average, which are synthetic workloads that do not involve actual training.} We assume that there are sufficient tasks so that when a block is generated, all $g_p$ provers in a group can be immediately assigned new tasks. To evenly distribute miners over time with a bootstrap (i.e., all miners join the network in the genesis block), the actual ML training epochs vary over $[0.9\overline{E},1.1\overline{E}]$. The simulation is run in a stage-based manner, which synchronizes all miners. In each stage, a miner trains the ML model for $\tau=4$ epochs and obtains a BGO, which is realized by coin tosses with success probability $p$ in the simulation. We assume that a miner takes $8$ epochs ($2$ stages) to download a dataset, and it takes $4$ epochs ($1$ stage) to upload and download the model weight. Adding the $1$-stage verification time, the total latency to verify a block is $2$ stages for the appointed committee of verifiers, and $4$ stages for the rest of the network (since the appointed committee downloads the dataset earlier). We run simulations long enough so that the miners are fully scattered and that our results are not affected by the bootstrap.

\begin{figure}[t]
	\centering
	\includegraphics[width=\columnwidth]{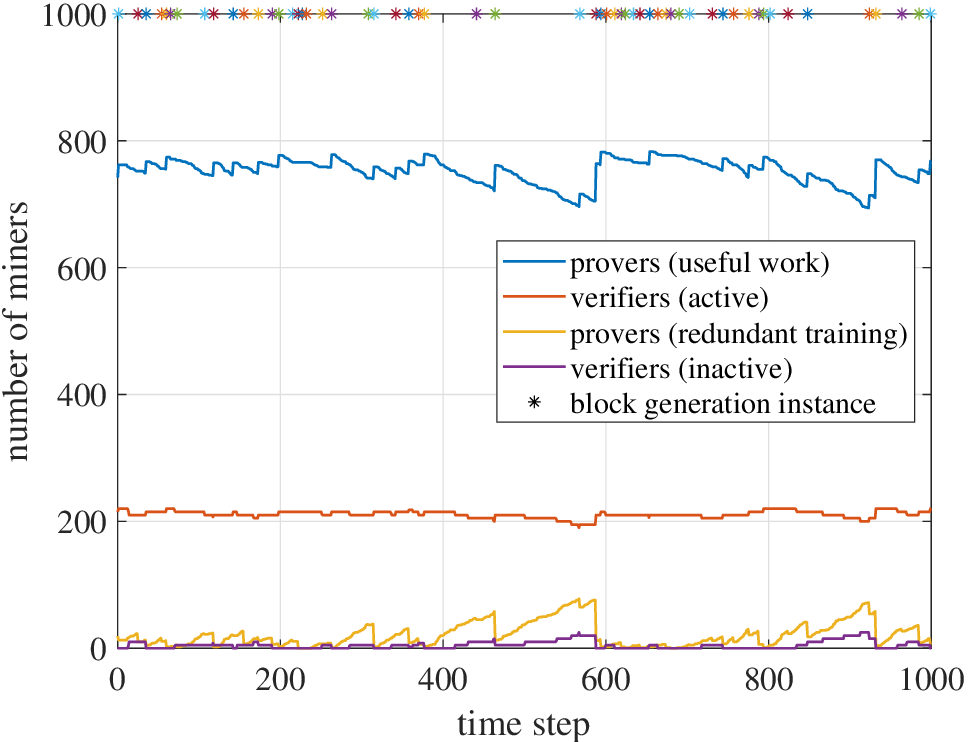}
	\caption{Evolution of provers and verifiers over time.}
	\label{fig:fig_sim1}
\end{figure}

\textcolor{black}{We first set $p=0.0001$, $\alpha=10$, and the simulation prototype runs smoothly with an average block generation interval of $17.3$ stages and a fork rate of $0.04$. The simulation prototype with the above parameters can be mapped to real-life ML training in the following manner. Suppose the ML task is to train a small convolutional neural network (roughly 10MB) on the MNIST dataset (roughly 20MB), which takes about 0.5s per epoch on a single GPU. There are 1000 miners with a bandwidth of 5MB/s each, so it takes about 4 seconds (8 epochs) to download the dataset and 2 seconds (4 epochs) to download a model. Therefore, the average block generation is around 35 seconds, which is between Ethereum and Litecoin, and the fork rate is consistent with short-interval blockchain systems.}

In Fig.~\ref{fig:fig_sim1}, the evolution of provers and verifiers as well as the block generation instances are presented, where we run simulations long enough and select a later reference point as the starting point for plotting the results so that the results represent a steady state situation. One can see that, the number of provers doing useful work (i.e., training an ML task) fluctuates slightly below $800$, while the number of active verifiers is slightly above $200$. In each stage, some provers finish their task, submit a new SD, and wait to be reappointed, leading to a decrease in the number of useful work provers. To maintain consistent mining power, these provers switch to redundant training (i.e., use a previous model weight to generate the template block). {\color{black}For example,} during time-steps $400-600$, when the block generation interval increases, the number of provers doing useful work drops more rapidly and the number of provers doing redundant work increases accordingly. On the other hand, since verifiers can only start task verification after the provers finish their task, a verifier typically takes a longer time to finish its duty than a prover. Therefore, as the system stabilizes, there are slightly more verifiers than $ng_v/g=200$. \textcolor{black}{Overall,} Fig.~\ref{fig:fig_sim1} shows that our framework works well under well-designed system parameters.

\begin{figure}[t]
	\centering
	\includegraphics[width=\columnwidth]{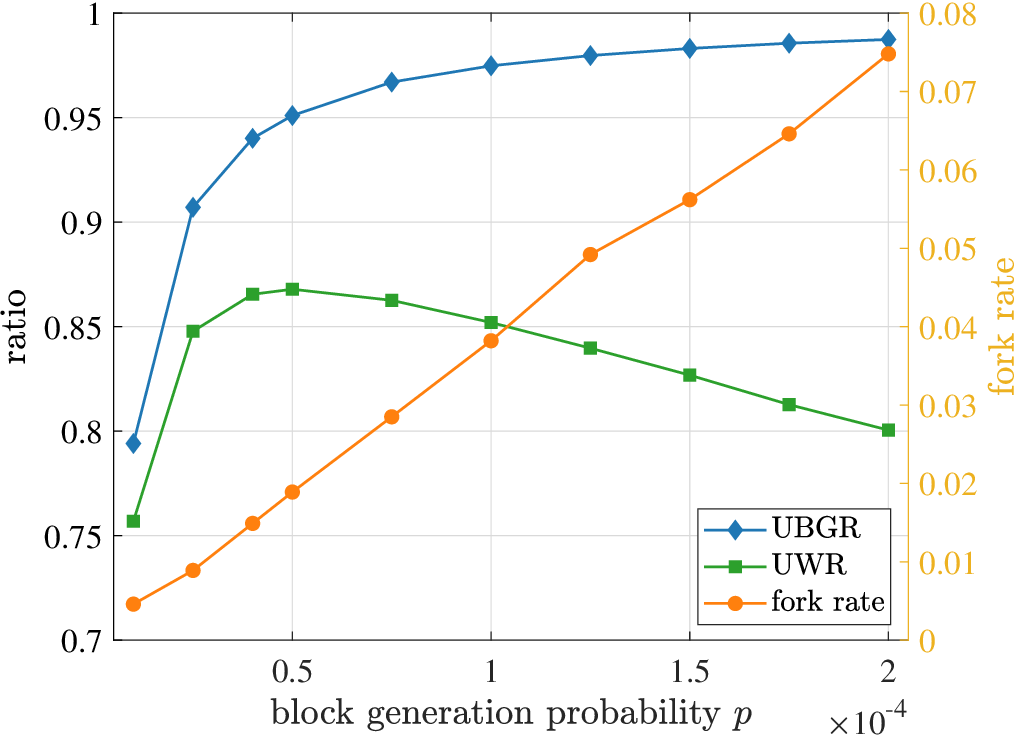}
	\caption{Useful block generation ratio, useful work ratio and fork rate under variant block generation probabilities $p$.}
	\label{fig:fig_sim2}
\end{figure}

Fig.~\ref{fig:fig_sim2} shows the useful block generation ratio, useful work ratio, and the fork rate under varying block generation probabilities $p$. Note that $p$ is chosen based on the amount of miners in the system. Both the useful block generation ratio and the fork rate grow with the increase of $p$, while the useful work ratio grows at first then decreases. This is because a higher $p$ reduces the block generation intervals and the time a prover spends on redundant training before its SD is recorded on-chain to get a new task. However, this also means that forks are more likely to occur, and more block verification is needed by the entire blockchain network, decreasing the UWR. In our simulation systems with $n=1000$ miners, the UWR peaks at $p=0.00005$, where about $86.8\%$ of all the computations in the system are useful in the sense that they involve training a new ML stage contributing to the final models that has not been done before by anyone in the system, and the fork rate is not too high ($1.9\%$). This indicates that, though computational overheads are inevitable, we can fine-tune system parameters to guarantee that the useful computations in the blockchain system are still sufficiently efficient. \textcolor{black}{Notably, for competition-based PoLs, since only the winner's block is useful and contributes to blockchain security, both the UBGR and the UWR are upper bounded by the computing power share of the largest miner in the network, which, according to the security assumptions, is strictly smaller than $1/2$ and creates a trade-off between useful work and decentralization. }

\begin{figure}[t]
	\centering
	\includegraphics[width=\columnwidth]{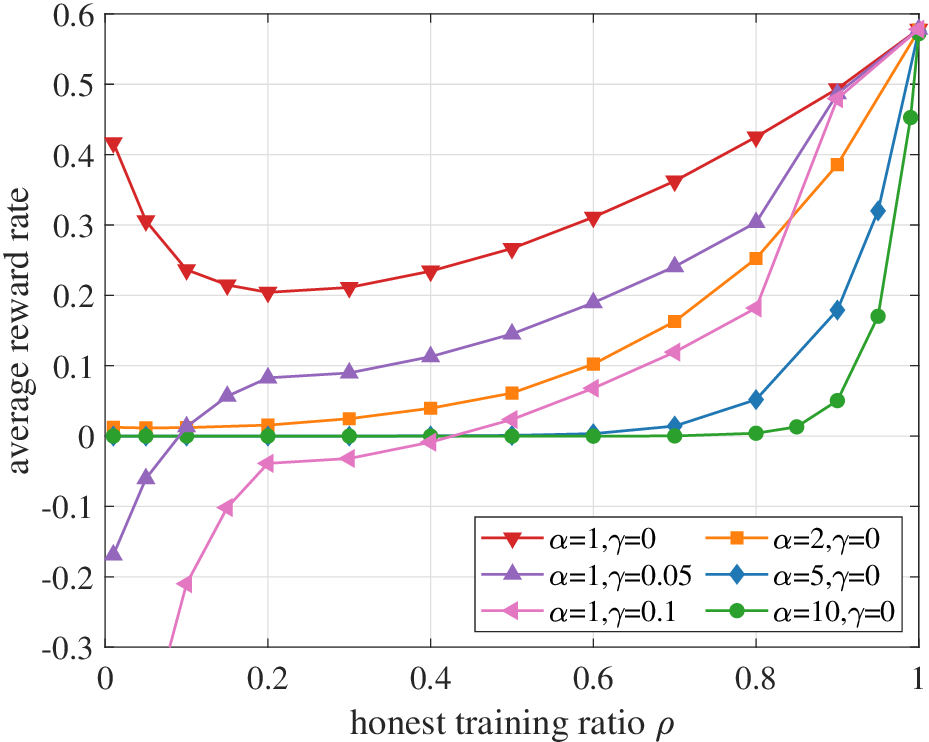}
	\caption{Average reward rate under different honest training ratio $\rho$.}
	\label{fig:fig_sim3}
\end{figure}

In Fig.~\ref{fig:fig_sim3}, we fix $p=0.0001$, $g_v=5$ and display the task reward rate of miners using different honest training ratio, $\rho$, given varying number of stages verified by each verifier, $\alpha$, and penalty-to-task-reward ratio, $\gamma$. The task reward is $1$ per training stage, thus $\overline{R_t}=1000$ and the penalty is $\gamma\overline{R_t}$ for each task trained dishonestly and caught. We assume the CTF is used so a dishonestly trained stage is caught with probability $\frac{1}{2}$. We see that, when $\alpha=1$ and there is no penalty, as $\rho$ rises, the average reward ratio first decreases then increases. In fact, when $\rho=0$, theoretically the dishonest prover can obtain infinite reward by trying infinite times since there is no penalty, which exceeds honest training ($\rho=1$). However, it takes time for a new SD to be uploaded, limiting the amount of times the dishonest prover can try. Moreover, as $\alpha$ increases, the probability that the dishonest prover passes verification drops exponentially, such that it can hardly get any reward even without penalty. When $\gamma$ increases, even for $\alpha=1$, the dishonest prover's reward ratio drops significantly as $\rho$ decreases. In fact, from \eqref{eq:lb-ctf}, when the committee verifies $g_v\alpha=5$ stages per task, $\gamma=0.05$ is sufficient to incentivize honesty.

\section{Conclusion} \label{sec:conclusion}
In this paper, we proposed SEDULity, a novel PoL framework that not only trains ML models efficiently, but also maintains the blockchain integrity in a distributed and secure manner. We presented the PoL framework design, which combines ML with block generation and builds a useful function on top of the ML model training. We discussed why our framework is indeed distributed, secure and efficient, and showed that the key intuition can be extended to other types of useful work. We designed an incentive mechanism, which slightly adjusts the proposed framework to incentivize task verification. We theoretically derived a sufficient honest condition such that provers are incentivized to train fully honestly, which guarantees the usefulness of the framework. Simulation results validated the feasibility and performance of the proposed consensus mechanism. 

\bibliographystyle{IEEEtran}
\bibliography{reference}

\end{document}